\documentclass[%
reprint,
superscriptaddress,
showpacs,preprintnumbers,
 amsmath,amssymb,
aps,
pre,
floatfix
]{revtex4-1}

\usepackage{float}
\usepackage{graphicx}
\usepackage{dcolumn}
\usepackage{bm}
\usepackage{bbold}
\usepackage[multidot]{grffile}
\usepackage[colorlinks=true,allcolors=blue, breaklinks=true]{hyperref}
\newcommand{\vect}[1]{\bm{\mathrm{#1}}}

\DeclareMathOperator{\dd}{d\!} 
\DeclareMathOperator{\tr}{tr}

\usepackage{xcolor}

\usepackage{xspace}
  
\begin{document}


\title{Desynchronization induced by time-varying network}

\author{Maxime Lucas}
\email{m.lucas@lancaster.ac.uk}
\affiliation{Department of Physics, Lancaster University, Lancaster LA1 4YB, United Kingdom}
\affiliation{Dipartimento di Fisica e Astronomia, Universit{\`a} di Firenze, INFN and CSDC, Via Sansone 1, 50019 Sesto Fiorentino, Firenze, Italy}
\author{Duccio Fanelli}
\email{duccio.fanelli@gmail.com}
\affiliation{Dipartimento di Fisica e Astronomia, Universit{\`a} di Firenze, INFN and CSDC, Via Sansone 1, 50019 Sesto Fiorentino, Firenze, Italy}
\author{Timoteo Carletti}
\email{timoteo.carletti@unamur.be}
\affiliation{naXys, Namur Institute for Complex Systems, University of Namur, B5000 Namur, Belgium}
\author{Julien Petit}
\email{julien.petit@student.unamur.be}
\affiliation{naXys, Namur Institute for Complex Systems, University of Namur, B5000 Namur, Belgium}
\affiliation{Department of Mathematics, Royal Military Academy, B1000 Brussels, Belgium}

\date{\today}

\begin{abstract}

The synchronous dynamics of an array of excitable oscillators, coupled via a generic graph, is studied. Non homogeneous perturbations can grow and destroy synchrony, via a self-consistent instability which is solely instigated by the intrinsic network dynamics. By acting on the characteristic time-scale of the network modulation, one can make the examined system to behave as its (partially) averaged analog. This result if formally obtained by proving an extended version of the averaging theorem, which allows for partial averages to be carried out. As a byproduct of the analysis, oscillation death are reported to follow the onset of the network driven instability.  
\end{abstract}

\maketitle

Natural and artificial systems are often composed of individual oscillatory units, coupled together so as to yield complex collective dynamics \cite{nicolis1977, kuramoto1984,goldbeter1997, pikovsky2003, strogatz2004}. Weak coupling of non-linear oscillators leads to synchronization \cite{pikovsky2003}, a condition of utmost coordination which is eventually met when the parts of a system operate in unison. Synchronization has been addressed theoretically in a 
wide range of settings, climbing the hierarchy of complexity from simple unidirectionally phase forced oscillator, with a fixed frequency~\cite{pikovsky2003} and more recently a time-varying frequency~\cite{suprunenko2013, lucas2017}, to large populations of mutually interacting, individually oscillating, entities \cite{kuramoto1984}. The simultaneous flashing of fireflies and the rhythmic applause in a large audience are representative examples both ascribable to the vast and multifaceted realm of synchronization phenomena \cite{strogatz2004}. Synchronisation of self-sustained oscillators on complex networks has attracted considerable interest in the last decade, the emphasis being primarily placed on the pivotal role exerted by the topology of the graph that shapes the underlying couplings \cite{barahona2002, arenas2008}. Other studies elaborated on the effect produced by imposing an external perturbation, such as noise \cite{strogatz1991}, a (fixed-frequency) pacemaker forcing \cite{kori2006, ott2008}, or more recently an external modulation of frequencies \cite{petkoski2012, lancaster2016, pietras2016}. Delays in the network of couplings have been also enforced and their impact on the synchronizability property thoroughly assessed \cite{earl2003}. 

At the other extreme entirely, when the coupling strength is made to increase, oscillations may go extinct. Oscillation death is observed in particular when an initially synchronized state evolves torwards an asymptotic inhomogeneous steady configuration \cite{hou2003, zou2009, nakao2010, challenger2015}, in response to an externally injected  perturbation \cite{koseska2013}. Understanding the mechanisms that drive the suppression of the oscillations in spatially extended systems, assimilated to disordered networks, may prove to be relevant for e.g. neuroscience applications. The ability of disrupting synchronous oscillations could be in fact exploited as a dynamical regulator \cite{kim2005,kumar2008, asllani2017}, to oppose pathological neuronal states that are found to consistently emerge in Alzheimer and Parkinson diseases. Landscape fragmentation and then dispersal among connected patches sits at the origin of oscillation death in ecology, with its noteworthy fallout in terms of diversity and stability~\cite{arumugam2016}. To date, oscillation death has been mostly analyzed on static networks. In many cases of interests \cite{holme2012, holme2015, masuda2016}, however, links are intermittently active and signals can crawl only when connections are functioning: diseases spread through physical proximity, pathogens flowing therefore on dynamic contact graphs; neural and brain networks can be also represented as time-varying graphs, resources driven activation playing a role of paramount importance. 

The inherent ability of a network to adjust in time acts as a veritable non-autonomous drive. In a recent Letter~\cite{petit2017}, the process of pattern formation for a multispecies model anchored on a time-varying network was analyzed.  It was in particular shown that a homogeneous stable fixed point can turn unstable, upon injection of a non-homogeneous perturbation, via a symmetry breaking instability which is reminiscent of the Turing mechanism \cite{turing1952}, but solely instigated by the intrinsic network dynamics. Starting from these premises, the aim of this work is to extend the theory presented in \cite{petit2017} to the relevant setting where the unperturbed homogeneous solution typifies as collection of synchronized limit-cycles. In other words, we will set to analyze how the synchrony of a large population of non-linear, diffusively coupled  oscillators may be disrupted by network plasticity. Surprisingly, oscillation death can be induced by a piecewise constant time-varying network, also when synchrony is guaranteed on each isolated network snapshot. Our analysis provides a solid theoretical backup to the work of \cite{sugitani2014}, where the oscillation death phenomenon is numerically observed on fast time-varying networks.

Consider two different species living on a network that evolves over time and denote by  $x_i$ and $y_i$ their respective concentrations, as seen on node $i$. The structural properties of the (symmetric) network are stored in a time-varying $N \times N$ weighted adjacency matrix $A_{ij}(t)$. For the ease of calculation, we will hereafter assume $N$ constant. Introduce the Laplacian matrix $\mathbf{L}$ whose elements read $L_{ij}(t) = A_{ij}(t) - K_i(t) \delta_{ij}$, where $K_i(t) = \sum_j A_{ij}(t)$ stands for the connectivity of node $i$, at time $t$. The coupled dynamics of $x_i$ and $y_i$, for $i=1,...,N$, is assumed to be ruled by the following, rather general equations: 

\begin{equation}
\begin{aligned}
\dot x_i &= f(x_i, y_i) + D_x \sum_{j=1}^N L_{ij}(t / \epsilon) x_j, \\
\dot y_i &= g(x_i, y_i) + D_y \sum_{j=1}^N L_{ij}(t / \epsilon) y_j, \\
\end{aligned}
\label{eq:net_i}
\end{equation}
where $D_x$ and $D_y$ are appropriate coupling parameters. Here, $f$ and $g$ are non-linear reaction terms, chosen in such a way that system (\ref{eq:net_i}) 
exhibits a homogeneous stable solution $(x_i,y_i) \equiv (\bar{x}(t),\bar{y}(t))$ $\forall i$ which is periodic of period $T$.  To state it differently, when $D_x=D_y=0$, the above system is equivalent to $N$ identical replica of a two dimensional deterministic model, which displays a stable limit-cycle. The homogeneous time-dependent solution obtained for $D_x \ne 0 \ne D_y$, 
when setting in phase the self-sustained oscillations on each node of the collection, corresponds to the synchronized regime that we shall be probing in the forthcoming 
investigation. The parameter $\epsilon$ controls the time-scale of the Laplacian dynamics.  We will specifically inspect the case of a network that is periodically rearranged in time and denote with $T_s$ the period of the network modulation, as obtained for  $\epsilon=1$.  By operating in this context, we will show that synchronization can be eventually lost when forcing $\epsilon$  below a critical threshold.  When successive swaps between two static network configurations are considered over one period $T_s$ (as it is the case, in the example addressed in the second part of the paper), $\epsilon$ sets the frequency of the blinking.  The extension to non-periodic settings is straightforward,  as discussed in details in \cite{petit2017}.

To proceed with the analysis we compactify the notation by introducing the $2N$-element vector $\vect{x} = (x_1, \dots, x_N, y_1, \dots, y_N)^T$. The dynamics of the system can be 
hence cast in the form:
\begin{equation}
\dot{\vect{x}} = \mathcal{F}(\vect{x}) + \mathcal{L}(t / \epsilon) \vect{x} ,
\label{eq:x_net}
\end{equation}
where $\mathcal{F}(\vect{x}) = (f(x_1,y_1), \dots, f(x_N, y_N), g(x_1, y_1), \dots,\allowbreak g(x_N, y_N))^T$; the $2N \times 2N$ block diagonal matrix $\mathcal{L}$ reads:
\begin{equation}
\mathcal{L}(t) = 
\left(
\begin{array}{cc}
D_x \, \vect{L}(t) & 0 \\
0 & D_y \, \vect{L}(t)
\end{array}
\right).
\end{equation} 

As mentioned above, the non-linear reaction terms, now stored in matrix $\mathcal{F}$, are chosen so as to have a stable limit-cycle in the uncoupled setting $D_x = D_y = 0$.  The stability of the limit-cycle $(\bar{x}(t),\bar{y}(t))$ can be assessed by means of a straightforward application of the Floquet theory.  To this end, we focus on the two dimensional system obtained in the uncoupled limit and introduce a perturbation of the time-dependent equilibrium, namely $\delta \vect{x} = (x - \bar{x},  y - \bar{y})^T$. Linearizing the governing equation yields $\delta \dot{\vect{x}} = \mathcal{J}(t) \delta \vect{x}$, where $\mathcal{J}(t) = \partial_{\vect{x}} \mathcal{F}(\vect{x})$ is periodic of period $T$.  Let us label with $\Phi(t)$ 
a fundamental matrix of the system.  Then, for all 
$t$, there exists a non-singular, constant matrix  $\vect{B}$ such that: 
\begin{equation}
\Phi(t + T) = \Phi(t) \vect{B}.
\label{eq:floquet_fund}
\end{equation}
Moreover,  $\det \vect{B} = \exp \left[ \int_0^T \tr \mathcal{J}(t) \dd t \right]$. The matrix $\vect{B}$
depends in general on the choice of the fundamental matrix $\Phi(t)$. Its eigenvalues, $\rho_i$ 
with $i=1,2$, however, do not. These are called the Floquet multipliers and yield the Floquet exponents, defined as $\mu_i = T^{-1} \ln \rho_i$. 
Solutions of the examined linear system can then be written: 
\begin{equation}
\vect{x}(t) = a_1 \vect{p_1}(t) e^{\mu_1 t} + a_2 \vect{p_2}(t) e^{\mu_2 t},
\end{equation}
where the $\vect{p_i}(t)$ functions are $T$-periodic, and $a_i$ are constant coefficients set by the initial conditions. When the system is linearized about limit-cycles arising from first-order equations, one of the Floquet exponents is identically equal to zero, $\mu_1=0$. The latter is associated with perturbations along the longitudinal direction of the limit-cycle: these perturbations are neither amplified nor damped as the motion progresses. The other exponent, $\mu_2$ takes instead negative real values, if the limit-cycle is stable, meaning that perturbations in the transverse direction are bound to decay in time. 

We now turn to discussing the original system (\ref{eq:x_net}). The reaction parameters are set so to yield a stable limit-cycle for $D_x = D_y = 0$. Furthermore, we assume the oscillators to be initially synchronized, with no relative dephasing. We then apply a small, nonhomogeneous, hence node-dependent perturbation and set to explore the conditions which can yield a symmetry breaking instability of the synchronized regime, from which the  oscillation death phenomenon might eventually emerge. We are in particular 
interested in elaborating on the role played by the non-autonomous network dynamics in seeding the aforementioned instability. 
Introduce a  small inhomogeneous perturbation around the synchronous solution $\delta \vect{x} = (x_1 - \bar x, \dots, x_N - \bar x, y_1 - \bar y, \dots, y_N - \bar y)^T$, and linearize the governing equation \eqref{eq:x_net} so as to yield:

\begin{equation}
\delta \dot{\vect{x}} = [\mathcal{J}(t) +  \mathcal{L}(t / \epsilon)]  \delta \vect{x}.
\label{eq:dx_net_original}
\end{equation}

This is a non-autonomous equation, and it is difficult to treat it analytically \cite{kloeden2011}, owing in particular to the simultaneous presence of different periods. 
To overcome this limitation, and gain analytical insight into the problem under scrutiny, we introduce the averaged Laplacian $\langle  \mathcal{L} \rangle = 1/T_s \int_0^{T_s}  \mathcal{L} \dd t$ and define the following system: 
\begin{equation}
\dot{\vect{y}} = \mathcal{F}(\vect{y}) + \langle\mathcal{L} \rangle \vect{y} .
\label{eq:x_net_av}
\end{equation}
As we will rigorously show in the following, the stability of the synchronous solution of system (\ref{eq:x_net}) is eventually amenable to that of system (\ref{eq:x_net_av}). Stated differently, assume that an external non-homogeneous perturbation can trigger an instability in system (\ref{eq:x_net_av}). Then, $\epsilon^*$ exists such that the original system (\ref{eq:x_net}) is also unstable for $0 < \epsilon < \epsilon^*$. In other words, by tuning sufficiently small the parameter $\epsilon$, and thus forcing a high frequency modulation of the network Laplacian, one can yield a loss of stability of the synchronous solution. Oscillation death can eventually emerge as a possible stationary stable attractor of the ensuing dynamics, promoted by the inherent ability of the network to adjust in time. 

As a first step towards proving the results, we shall rescale time as $\tau = t / \epsilon$. Eq.~\eqref{eq:dx_net_original} can be hence cast in the equivalent form: 
\begin{equation}
\delta \vect{x}^{\prime} = \epsilon [(\mathcal{J}(\epsilon \tau) +  \mathcal{L}(\tau)] \delta \vect{x},
\label{eq:dx_tau}
\end{equation}
where the prime denotes the derivative with respect to the new time variable $\tau$.  The {\it partially averaged} version of  (\ref{eq:dx_tau}) [or alternatively the linear version of system \eqref{eq:x_net_av}, after time rescaling], reads
$\delta \vect{y}^{\prime} = \epsilon [\mathcal{J}(\epsilon \tau) +  \langle \mathcal{L} \rangle] \delta \vect{y}$. In the following, we will show that
$\delta \vect{y}(t) - \delta \vect{x}(t) \in \mathcal{O}(\epsilon)$ for up to a time $\tau \in \mathcal{O}(1/\epsilon)$, provided that $\delta \vect{y}(0) = \delta \vect{x}(0)$ and  for $\epsilon < \epsilon^*$. This conclusion builds on a theorem that we shall prove hereafter in its full generality, and which extends the realm of applicability of the usual averaging theorem. Denote $\vect{x} \in \mathbb{R}^\Omega$, and consider the following equation
\begin{equation}
\dot{\vect{x}} = \epsilon f_1(\vect{x}, \epsilon t) + \epsilon f_2(\vect{x}, t),
\label{eq:adapted_avg_start}
\end{equation}
where $f_1(\vect{x}, t)$ is $T$-periodic in $t$, and $f_2(\vect{x}, t)$ is $T_s$-periodic in $t$. Notice that $f_1(\vect{x}, \epsilon t)$  is $T/\epsilon$-periodic. It is assumed that $f_1$ and $f_2$ and their derivative are well behaved Lipschitz-continuous functions. Observe incidentally that Eq.~\eqref{eq:dx_tau} is recovered by replacing $t \mapsto \tau$, $\vect{x} \mapsto \delta \vect{x}$, $f_1(\vect{x}, \epsilon t) \mapsto \mathcal{J}(\epsilon \tau) \delta \vect{x}$,  $f_2(\vect{x}, t) \mapsto \mathcal{L}(\tau) \delta \vect{x}$ and $\Omega=2N$.

The standard version of the averaging theorem \cite{verhulst1990} requires dealing with periodic functions, whose periods are  independent of $\epsilon$. This is obviously not the case for $f_1(\cdot, \epsilon t)$.  To bypass this technical obstacle,  we will adapt the proof in \cite{verhulst1990} to yield an alternative formulation of the theorem which allows for partial averaging to be performed. Define: 

\begin{equation}
u(\vect{x}, t) = \int_0^t \dd s [f_2 (\vect{x}, s) - \langle f_2 \rangle],
\label{eq:u_def}
\end{equation}
where $\langle f_2 \rangle = 1/T_s \int_0^{T_s} f_2 (\vect{x}, t) \dd t$ is the average of $f_2$ over its period. Introduce then the {\it near-identity} transformation
\begin{equation}
\vect{x}(t) = \vect{z}(t) + \epsilon u(\vect{z}(t), t), 
\end{equation}
which yields
\begin{equation}
\dot{\vect{x}} = \dot{\vect{z}} + \epsilon \frac{\partial u}{\partial \vect{z}}  \dot{\vect{z}}  + \epsilon \frac{\partial u}{\partial t} .
\label{eq:x_dot2}
\end{equation}
Moreover, $\partial u / \partial t \, (\vect{z}, t) = f_2 (\vect{z}, t) - \langle f_2 \rangle $ by definition of  $u$, see Eq.~\eqref{eq:u_def}. Then making use of  Eq.~\eqref{eq:adapted_avg_start}, it is straightforward to get:
%
%
%
\begin{equation}
\begin{split}
\overbrace{\left[\mathbb{1}  + \epsilon \frac{\partial u}{\partial \vect{z}} \right]}^{\equiv \Gamma} \dot{\vect{z}} 
=& \epsilon \left[ f_1(\vect{z} + \epsilon u, \epsilon t) +  f_2 (\vect{z} + \epsilon u, t) - f_2 (\vect{z}, t) + \langle f_2 \rangle \right],\\
\end{split}
\label{eq:z_inv}
\end{equation}
Invoking the Lipschitz-continuity of $f_2$ and the boundedness of $u$ yields :
\begin{equation}
\begin{split}
|| f_2 (\vect{z} + \epsilon u, \epsilon t) - f_2 (\vect{z}, \epsilon t) || &\le L \epsilon || u (\vect{z}, \epsilon t) ||,\\
&\le L \epsilon M,
\end{split}
\end{equation}
where $L$ and $M$ are positive constants. Hence:

\begin{equation}
\begin{split}
\Gamma  \dot{\vect{z}} 
=& \epsilon f_1(\vect{z} + \epsilon u, \epsilon t) + \epsilon \langle f_2 \rangle + \mathcal{O}(\epsilon^2), \\
\simeq& \epsilon f_1(\vect{z}, \epsilon t) + \epsilon \langle f_2 \rangle.
\end{split}
\label{eq:z_inv}
\end{equation}

We do not know in general if $\Gamma$ is invertible, but the identity is and, by continuity, any matrix sufficiently close to it. Hence, there exists a critical value $\epsilon^* \ll 1$ such that $\Gamma$ is invertible, if $0<\epsilon < \epsilon^*$. We will return later on providing a self-consistent estimate for the critical threshold $\epsilon^*$. Up to order $\mathcal{O}(\epsilon)$, we have:
\begin{equation}
\Gamma^{-1} \simeq \left[ \mathbb{1}  - \epsilon \frac{\partial u}{\partial \vect{z}}   \right].
\end{equation}
Hence finally, 
\begin{equation}
\dot{\vect{z}} \simeq \epsilon [ f_1(\vect{z}, \epsilon t) + \langle f_2 \rangle ] .
\label{eq:z_dot}
\end{equation}
In conclusion, system \eqref{eq:adapted_avg_start} behaves like its {\it partially averaged} version \eqref{eq:z_dot}, for times which grow like $1/\epsilon$, when $\epsilon$ is made progressively smaller. Back to the examined model, system~\eqref{eq:dx_tau}  stays thus close to its partially averaged homologue: 
\begin{equation}
\delta \vect{y}^{\prime} = \epsilon [\mathcal{J}(\epsilon \tau) +  \langle \mathcal{L} \rangle ]  \delta \vect{y},
\end{equation}
which, in terms of the original time scale $t$ amounts to:
\begin{equation}
\delta \dot{\vect{y}} = \mathcal{M}(t) \delta \vect{y},
\label{eq:partially_avgd}
\end{equation}
where $\mathcal{M}(t) = \mathcal{J}(t) +  \langle \mathcal{L} \rangle$ is a $T$-periodic $2 N \times 2 N$ matrix. It is worth emphasising that systems  (\ref{eq:x_net}) 
and (\ref{eq:partially_avgd}) agree on times $\mathcal{O}(1)$, owing to the definition of the variable $\tau$. Imagine conditions are set so that an externally imposed, non-homogeneous perturbation may disrupt  the synchronous regime, as stemming from Eq.~(\ref{eq:partially_avgd}). Then,  the same holds when the perturbation is made to act on system (\ref{eq:x_net}), the factual target of our analysis.  The onset of instability of (\ref{eq:x_net}) can be hence rigorously assessed by direct inspection of its partially averaged counterpart~(\ref{eq:x_net_av}), which yields the linear problem (\ref{eq:partially_avgd}). Patterns established at late times can be however different,  the agreement between the two systems being solely established at short times. 

System (\ref{eq:partially_avgd}) can be conveniently studied by expanding the perturbation on the basis of the average Laplacian operator,  $\langle \vect{L} \rangle = 1/T_s \int_0^{T_s} \vect{L} \dd t$ \footnote{The diagonalizability of the Laplacian matrix is a minimal requirement for the analytical treatment to hold true. This condition is trivially met when the network of couplings is assumed symmetric, as in the example worked out in the following.}. Introduce $\vect{\phi}^{(\alpha)}$, such that $\sum_{j=1}^N \langle L \rangle_{ij} \phi^{(\alpha)}_j =  \Lambda_{\alpha} \phi^{(\alpha)}_i$, where $\Lambda_{\alpha}$ stands for the eigenvalues of $\langle \vect{L} \rangle$ and $\alpha=1, \dots, N$. Note that the eigenvectors are time-independent, as the averaged network (hence, the Laplacian) is. Write then $\delta x_i(t) = \sum_{\alpha=1}^N c^x_{\alpha}(t) \phi^{(\alpha)}_i$ and $\delta y_i(t) = \sum_{\alpha=1}^N c^y_{\alpha}(t) \phi^{(\alpha)}_i$, where 
$c^x_{\alpha}$ and $c^y_{\alpha}$ encode the time-evolution of the linear system~\cite{nakao2010, asllani2014, asllani2014b}. Plugging the above ansatz into equation~\eqref{eq:partially_avgd} yields the following consistency condition:
\begin{equation}
\dot{\vect{c}}_{\alpha} = \vect{M}_{\alpha} (t)  \vect{c}_{\alpha},
\label{eq:2d}
\end{equation}
where $\vect{c}_{\alpha} = (c^x_{\alpha} , c^y_{\alpha} ) ^T$, and $\vect{M}_{\alpha} (t) = 
\vect{J}(t) + \Lambda_{\alpha}
\left(
\begin{array}{cc}
D_x & 0\\
0 & D_y  \\
\end{array}
\right)
$. 
The fate of the perturbation is determined upon solving the above $2 \times 2$ linear system, for each  $\Lambda_{\alpha}$. To this end, remark that $\vect{M}_{\alpha}$  is periodic, with period $T$, $\forall \alpha$. Solving system (\ref{eq:2d}) amounts therefore to computing the Floquet exponents $\mu_1^{(\alpha)}$ and  $\mu_2^{(\alpha)}$, for $\alpha=1, \dots, N$. The dispersion relation is obtained by selecting the largest  real part of $\mu_i^{(\alpha)}$, $i=1,2$, at fixed $\alpha$~\cite{challenger2015}. For undirected networks ($A_{ij}=A_{ji}$),  the Laplacian is symmetric and the $\Lambda_{\alpha}$ are real and non-positive~\footnote{This condition needs to be relaxed when dealing with directed graphs. The general philosophy of the calculation remains however unchanged, at the price of some technical complication as discussed in \cite{asllani2014}.}. For $\alpha=1$ the largest Floquet multiplier is  zero, since the model displays in its a-spatial version ($\Lambda_{1}=0$) a stable limit-cycle. We then sort the indices $(\alpha)$ in decreasing order of the eigenvalues, so that the condition $0 = \Lambda_{1} \ge \Lambda_{2} \ge ... \ge \Lambda_{N}$ holds. If the dispersion relation is negative $\forall$ $\Lambda_{\alpha}$ with $\alpha>1$, the imposed perturbation fades away exponentially: the synchronous solution is therefore recovered, for both the average system (\ref{eq:partially_avgd}), and its original analogue, in light of the above analysis, and for all $\epsilon$. Conversely, if the dispersion relation takes positive values, even punctually, in correspondence of  specific  $\Lambda_{\alpha}$, belonging to its domain of definition,  then the perturbation grows exponentially in time, for $\epsilon$ smaller than a critical threshold. The  initial synchrony for the original system (\ref{eq:x_net}) is hence lost and patterns may emerge.

\begin{figure}[h]
	\centering
	\includegraphics[width=\linewidth]{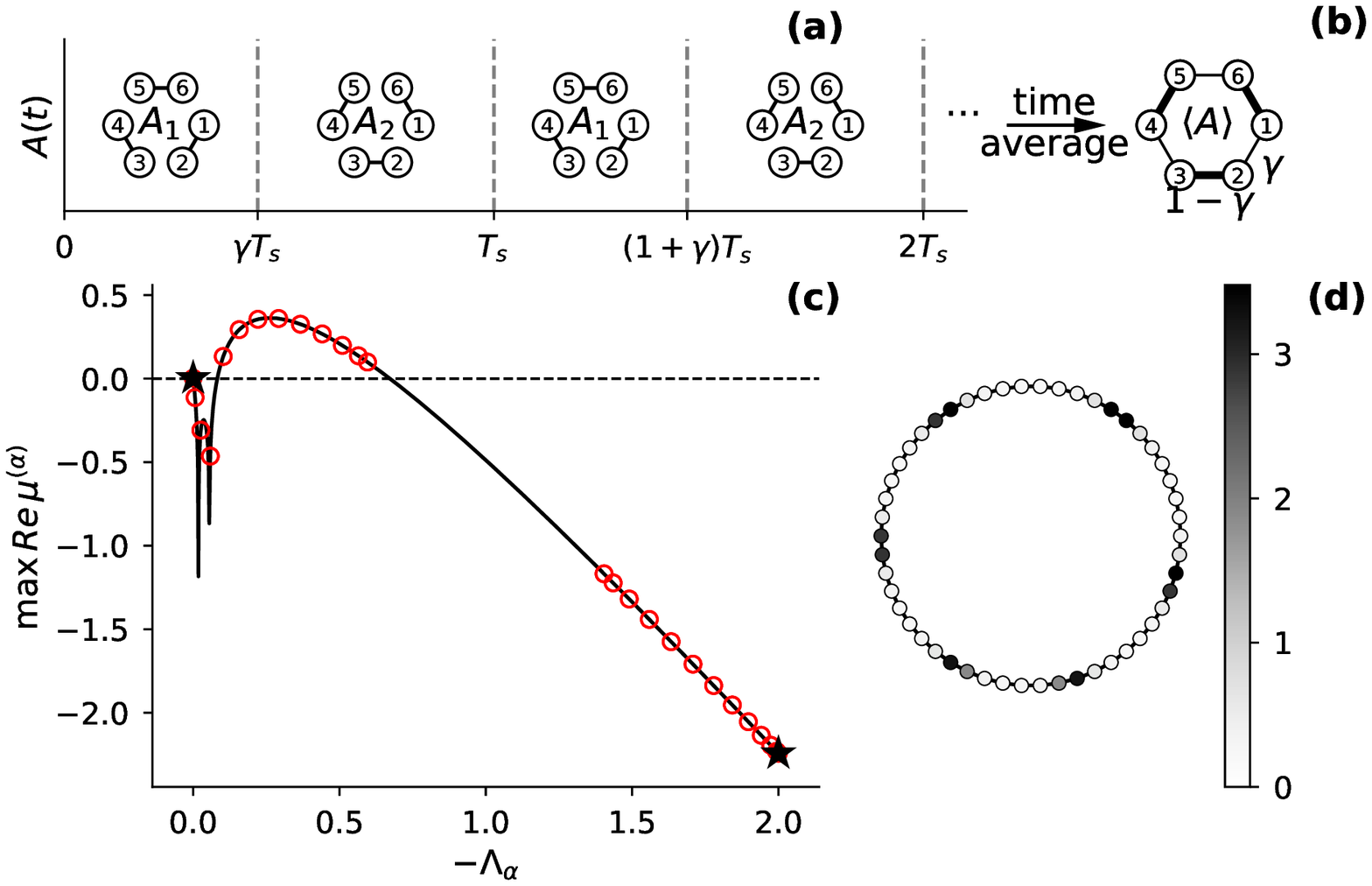}
	\caption{Instability in time-varying networks. (a) Dynamics of $\vect{A}(t)$, as obtained by alternating two static networks with adjacency matrices $\vect{A_1}$ and $\vect{A_2}$ (see main text for a detail account of the imposed couplings), over a cycle of time duration $T_s$. Each network in this illustrative example is made of $N = 6$ nodes. (b) The associated time-averaged network $\langle \vect{A} \rangle = \gamma \vect{A_1} + (1-\gamma) \vect{A_2}$. (c) Dispersion relation ($\max$ (Re $\mu^{\alpha})$ against $-\Lambda_{\alpha}$) obtained by assuming (i)  the averaged network $\langle \vect{A} \rangle$ (red circles), (ii) each static network (black stars) and (iii) the continuous support case (black curve). The networks are generated according to the procedure discussed in the main body of the paper, but assuming now $N = 50$. Other parameters are set to $b=4.5$, $c=2.5$, $D_x=2$, $D_y=20$, $T_s=1$, and $\gamma=0.3$. (d) Asymptotic, stationary stable patterns, obtained for $\epsilon = 0.1 < \epsilon^*$. Shades of grey represent the value of the $x$ variable.}
	\label{fig:dispersion_relation}
\end{figure}

To clarify the conclusion reached above, we shall hereafter consider a pedagogical example, borrowed from~\cite{petit2017}. Introduce the  
Brusselator model, a universally accepted theoretical playground for exploring the dynamics of autocatalytic reactions. This implies selecting $f(x,y) = 1 - (b+1) x + c x^2 y$ and 
$g(x,y) = b x - c x^2 y$, where $b$ and $c$ stand for free parameters.  For $b > c + 1$, the Brussellator model displays a limit-cycle. Following, \cite{petit2017} we then consider two networks, made of  an even number, $N$, of nodes arranged on a periodic ring, and label their associated adjacency matrices $\vect{A_1}$ and $\vect{A_2}$, respectively. Nodes are connected in pairs, via  symmetric edges. When it comes to the network encoded in $\vect{A_1}$, the couples are formed by the nodes labelled with the indexes $2k-1$ and $2k$ for $k=1,2,..., N/2$ [see panel (a) in Fig.~\ref{fig:dispersion_relation}]. The network specified via the adjacency matrix  $\vect{A_2}$ links nodes $2k$ and $2k+1$, with the addition of nodes $1$ and $N$ [as depicted in panel (a) in Fig.~\ref{fig:dispersion_relation}]. Both networks return an identical Laplacian spectrum, namely two degenerate eigenvalues $\Lambda_1=0$ and $\Lambda_N=-2$, with multiplicity $N/2$. The parameters of the Brussellator are set so that the synchronized solution is stable on each network, taken independently. This is illustrated in panel (c) of Fig.~\ref{fig:dispersion_relation}, where the corresponding dispersion relation (largest real part of the Floquet multipliers vs.  $-\Lambda_{\alpha}$) is plotted with black star symbols. 
Introduce now the time-varying network, specified by the  adjacency matrix $\vect{A}(t)$, defined as: 
\begin{equation}
\vect{A}(t) = 
\left\{
\begin{array}{ll}
 \vect{A_1} & \text{if} \mod(t, T_s) \in [0, \gamma[, \\
 \vect{A_2} & \text{if} \mod(t, T_s) \in [\gamma, 1[, \\
\end{array}
\right.
\label{eq:a(t)}
\end{equation}
where $\gamma$ (resp. $1-\gamma$) is the fraction of $T_s$ that the network spends in the configuration  specified by the adjacency matrix $\vect{A_1}$ (resp.  $\vect{A_2}$).
The average network is hence characterized by the adjacency matrix $\langle \vect{A} \rangle = \gamma \vect{A_1} + (1 - \gamma) \vect{A_2}$, see panel (b) in Fig.~\ref{fig:dispersion_relation}. We then set to consider the stability of the synchronized state in presence of a time-varying network, and resort to its static, averaged counterpart.  
The average network Laplacian has many more distinct eigenvalues, and these latter fall in a region where the largest real part of the Floquet exponents is positive, as can be appreciated in Fig.~\ref{fig:dispersion_relation}, panel (c), thus signaling the instability. The solid line stands for the dispersion relation that is eventually recovered when the couplings among oscillators extends on a continuum support and the algebraic Laplacian is replaced by the usual second order differential operator \cite{nakao2010, challenger2015}.  Since the dynamics hosted on the average network is unstable, the synchrony of the  homogenous state can be broken on the time-varying setting, by properly modulating $\epsilon$ below a critical threshold. This amounts in turn to imposing a fast switching between the two network snapshots, as introduced above. In Fig.~\ref{fig:dispersion_relation}, panel (d), the asymptotic pattern as displayed on a time-varying network, for a sufficiently small  $\epsilon$ is depicted. The nodes of the network are colored with an appropriate code chosen so as to reflect the asymptotic value of the density displayed by the activator species $x$. A clear pattern is observed which testifies on the heterogeneous nature of the density distribution, following the symmetry breaking instability seeded by the inherent network dynamics. Interestingly, the equilibrium density, as displayed on each node of the collection, converges to a constant:  synchronous oscillations,  which define the initial homogeneous state, are self-consistently damped to yield a stationary stable, heterogeneous distribution of the concentrations. This is the oscillation death phenomenon to which we made reference above. For the sake of clarity, this effect has been here illustrated with reference to a specific case study, engineered so as to allow for an immediate understanding of the key mechanism. The result reported holds however in general and apply to other realms of investigation where  time-varying network topology and  non-linear reaction terms are complexified at will.  

\begin{figure}[h]
\centering
\includegraphics[width=\linewidth]{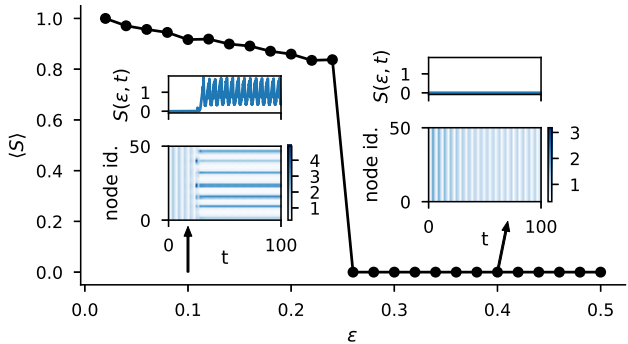}
\caption{Critical threshold $\epsilon^*$. Average pattern amplitude, $\langle S \rangle$, as a function of $\epsilon$, normalized to the amplitude of the pattern referred to the averaged network $\langle \vect{A} \rangle = \gamma \vect{A_1} + (1-\gamma) \vect{A_2}$ (as formally recovered in the limit $\epsilon \rightarrow 0$). Here, $N = 50$ nodes. Other parameters are set to $b=4.5$, $c=2.5$, $D_x=2$, $D_y=20$, and $\gamma=0.3$. (Insets) Dynamics of $x$, on each node, over time. Shades of blue represent the value of $x$. (Left) for $\epsilon = 0.1$, the synchronous solution is unstable. After a transient time, oscillation death is seen, and a heterogeneous pattern develops. (Right) for $\epsilon = 0.4$, the synchronous solution is stable. $S(\epsilon, t )$ is also plotted vs. time for the two considered settings. }
\label{fig:critical_epsilon}
\end{figure}

To shed further light onto the dynamics of the system, we introduce the macroscopic indicator
\begin{equation}
S(\epsilon, t) = \frac{1}{N}  \| \vect{x}(t) - \bar{\vect{x}}(t) \| ^ 2,
\label{S}
\end{equation}
where $\bar{\vect{x}}(t) = (\bar{x}, \dots, \bar x, \bar{y}, \dots, \bar y)$. $S(\epsilon, t)$ enables us to quantify the, time-dependent, cumulative deviation between individual oscillators trajectories, and the homogeneous synchronized solution. $S(\epsilon, t)$ will rapidly converge to zero, if the synchronous state is stable. Conversely, it will take non zero, positive values, when the imposed perturbation destroys the initial synchrony. To favor an immediate reading of the output quantities, we set to measure $\langle S \rangle$, the average of $S(\epsilon, t)$, on one cycle $T_s$. In formulae, $\langle S \rangle = T_s^{-1} \int_{t}^{t+T_s} S(\epsilon, u) \dd u$, where $t$ is larger than the typical relaxation time (transient). In Fig.~\ref{fig:critical_epsilon}, main panel, $\langle S \rangle$ is plotted against $\epsilon$, normalized to the value it takes in the limit $\epsilon \to 0$, for a choice of the parameter that corresponds to the dispersion relation depicted in Fig.~\ref{fig:dispersion_relation}. A clear, almost abrupt, transition is seen, for $\epsilon^* \simeq 0.25$, in qualitative agreement with the above discussed scenario. For $\epsilon < \epsilon^*$, the oscillation are turned into a stationary stable pattern, as displayed in the annexed panel. By monitoring $S(\epsilon, t)$ for a choice of $\epsilon$ below the critical threshold, one observes regular oscillations that can be traced back to the term $\bar{\vect{x}}$ in equation~(\ref{S}). At variance, synchronous oscillations prove robust  to external perturbation when $\epsilon > \epsilon^*$: the order parameter $S(\epsilon, t)$ is identically equal to zero, the  two contributions in the argument of the sum on the right hand side of equation (\ref{S}) canceling mutually. 
 
To conclude the analysis, we will provide an approximate theoretical estimate of the critical threshold $\epsilon$. The proof of the partial averaging theorem, as outlined above, assumes an invertible change of coordinates. It is therefore reasonable to quantify $\epsilon^*$ by determining the range $\epsilon$ for which the invertibility condition is matched~\cite{petit2017}. In formulae:
\begin{equation}
    \epsilon^* = \min\{\epsilon > 0 : \det \Gamma(\epsilon)=0 \} .
\end{equation}
Using the block structure of $\partial u / \partial \vect{z} = \int_0^{\tau} [\mathcal{L}(t) - \langle \mathcal{L} \rangle] \dd t$, one gets a more explicit form of the determinant 
\begin{equation}
\begin{split}
\det (\mathbb{1}_{2N} +& \epsilon \partial u / \partial \vect{z}) 
=\\
 & \det \left( \mathbb{1}_N + \epsilon D_x \int_0^{\tau} [\vect{L}(t) - \langle \vect{L} \rangle] \dd t \right) \\
 &\times 
\det \left( \mathbb{1}_N + \epsilon D_y \int_0^{\tau} [\vect{L}(t) - \langle \vect{L} \rangle] \dd t
\right) ,
\end{split}
\end{equation}
which is zero if either of the determinants is zero. A straightforward manipulation yields, for the inspected network model, the following closed expression:
\begin{equation}
\epsilon^* \simeq \frac{1}{\Lambda_{12}^N \gamma (1- \gamma) T} \min \left[ \frac{1}{D_x}, \frac{1}{D_y} \right], 
\end{equation}
where $\Lambda_{12}^N$ stands for the maximum eigenvalue (in absolute magnitude) of the operator $(\vect{L_1} - \vect{L_2})$, with $\vect{L_1}$ and $\vect{L_2}$ being the Laplacian matrices associated to the static networks as specified by the adjacency matrices $\vect{A_1}$ and $\vect{A_2}$. Performing the calculation returns $\epsilon^*=0.12$,  a coarse approximation of the exact critical value, as determined via direct numerical integration \footnote{As an alternative for computing $\epsilon^*$,  assume $T$ and $\epsilon T_s$ are commensurable (if not,  adjust the value of $\epsilon$ correspondigly) and define the common period for the reaction and diffusion parts, $$T_c = \mathrm{LCM}(T,\epsilon T_s).$$ Compute the Floquet multipliers for the $2N \times 2N$ system which is periodic with period $T_c$. Repeating the above procedure for decreasing values of $\epsilon$ (and making sure $T$ and $\epsilon T_s$ are still commensurable) yields the critical $\epsilon$, i.e. the largest $\epsilon$ for which not all $\mu_i$'s are negative.}. 

\begin{figure}[h]
\centering
\includegraphics[width=\linewidth]{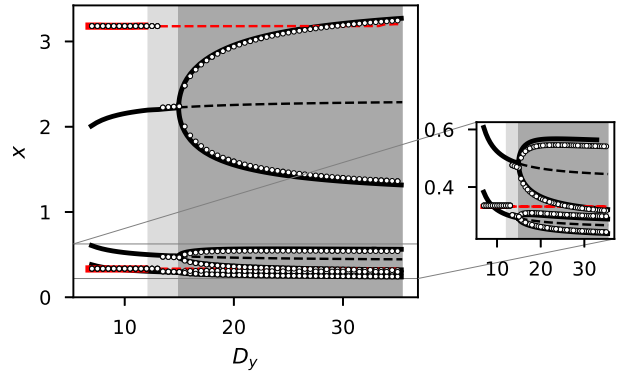}
\caption{Phase diagram for the Brusselator model coupled via time dependent pairwise exchanges, as illustrated in the caption of Fig.~\ref{fig:dispersion_relation}, with $N=6$. The equilibrium solutions relative to species $x$  are plotted by varying $D_y$, at fixed $D_x=2$. The stability is computed for the average analogue (\ref{eq:x_net_av}) of model (\ref{eq:net_i}). The horizontal (red, straight) lines  refer to the limit cycle:  the maximum and minimum values as attained by the uncoupled oscillators, over one period, are respectively displayed. Black lines stand for the  fixed points. Dashed lines refer to the unstable solutions, whereas solid lines implies stability. White circles are obtained from direct simulations of model (\ref{eq:net_i}) with $\epsilon=0.1$ and illustrate the oscillation death phenomenon, as discussed in the main text. The panel on the right is a zoom of the lower portion of the main plot. The shaded regions are drawn to guide the reader's eye across the different regimes: synchronization, oscillation death with 3-fixed point pattern, and oscillation death with 6-fixed point pattern correspond to the region in white, light gray, and dark gray, respectively. Notice that we chose to display a partial subset of the complete phase diagram. All stable manifolds are plotted. A limited subset of the existing unstable branches is instead shown for graphic requirements.}
\label{fig:bifurc}
\end{figure}

Finally, we shall inspect how the oscillation death phenomenon is influenced by the strength of the imposed coupling, here exemplified by the constant $D_y$, which we modulate when freezing $D_x$ to a nominal value. In Fig.~\ref{fig:bifurc} different attractors, and their associated stability, are depicted, for species $x$, for distinct choices of the control parameter $D_y$. 
Here, the Brussellator model is assumed as the reference reaction scheme; the network of pairwise exchanges ($N=6$), as illustrated in the caption of Fig.~\ref{fig:dispersion_relation}, is employed. The horizontal straight (red) lines refer to the limit cycle solution, and identify respectively the maximum and minimum value, as attained by the uncoupled oscillators, over one period. The solid trait marks the stable branch, while the dashed line is associated to the unstable solution. The bifurcation point is calculated analytically, from a linear stability analysis carried out for the average system (\ref{eq:x_net_av}).  Beyond the transition point, when the homogeneous solution breaks apart, three stable solutions are shown to exist, corresponding to distinct values of the concentration $x$.  These latter branches protrude inside the region where synchronous oscillations are predicted to be stable: the unstable manifolds which delineate the boundaries of the associated basins of attractions are not displayed for graphic requirements. Open (white) circles follow direct integration of model (\ref{eq:net_i}). In the simulations, $\epsilon$ is set to 0.1: the slight discrepancy between predicted and observed value of $D_y$ (at the onset of the desynchronization) stems from finite size corrections (the theory formally applies to the idealized setting $\epsilon \rightarrow 0$). When synchrony is lost,  the system evolves towards an asymptotic state that displays oscillation death: each node is associated to a stationary stable density, which is correctly explained by resorting to the average model approximation (\ref{eq:x_net_av}). Increasing further the coupling strength  $D_y$, results in a significant complexification of the phase space diagram, which considerably enrich the zoology of the emerging oscillation death patterns,  as displayed in Fig.~\ref{fig:bifurc} above the supercritical pitchfork bifurcations.

Summing up, we have here considered the synchronous dynamics of a collection of self-excitable oscillators, coupled via a generic graph. The plasticity of the underlying network of couplings, i.e. its inherent ability to adjust in time, may seed an instability which destroys synchrony.  The system endowed with a time-varying network of interlinked connections, behaves 
as its (partially) averaged analogue, provided the network dynamics is sufficiently fast. This result is formally established by resorting to  an extended version of the celebrated averaging theorem, which allows for partial averages to be performed. Interestingly, the network driven instability materializes in asymptotic, stationary stable patterns. These latter are to be regarded as a novel evidence for the oscillation death phenomenon.

This work has been funded by the EU as Horizon 2020 research and innovation programme under the Marie Sk\l{}odowska-Curie grant agreement No 642563.

\bibliographystyle{apsrev4-1}

\bibliography{bib}

\end{document}